\documentstyle[twocolumn,seceq]{jpsj}

\def\runtitle{Numerical Latent Heat Observation of the $q = 5$
Potts Model}
\def\runauthor{Tomotoshi {\sc Nishino} and Kouichi
{\sc Okunishi}}

\title{Numerical Latent Heat Observation of the $q = 5$ Potts
Model}

\author
{ Tomotoshi {\sc Nishino}\footnote{e-mail:
nishino@phys560.phys.kobe-u.ac.jp} and
Kouichi {\sc Okunishi}$^{1,}$\footnote{e-mail:
okunishi@godzilla.phys.sci.osaka-u.ac.jp} }

\inst
{ Department of Physics , Faculty of Science,
Kobe University, Rokkodai 657\\
$^1$Department of Physics, Graduate School of Science,
Osaka University, Toyonaka 560 }

\kword
{ Weak First-Order Transition, Latent Heat,
   Corner Transfer Matrix, Renormalization Group.}

\begin{document}
\sloppy
\maketitle

The $q$ state Potts model is one of the most intensively
studied classical two-dimensional lattice
models.~\cite{Potts,Wu} The model shows second-order phase
transition when $q \le 4$, and shows first-order transition when
$q > 4$.~\cite{Baxter,Wu} The critical  phenomena of the case $q
\le 4$ have been classified by the conformal  field
theory.~\cite{BPZ,FQS} Since critical exponents are exactly
known for $q = 2, 3$, and $4$, the Potts model has been used as
a reference system for numerical finite size
scaling~\cite{Fi,Ba} (FSS) analyses. Recently Carlon and
Igl\'oi~\cite{Carlon3} showed that the density matrix
renormalization group (DMRG),~\cite{Wh1,Wh2,Wh3} which has been
applied to classical lattice
models,~\cite{Nishino,Honda,Carlon1,Carlon2} is applicable to
the precise estimation of the critical exponents. They also
suggest the applicability of DMRG to the first-order phase
transition.~\cite{Carlon3,Carlon4}

The phase transition of the two-dimensional $q = 5$ Potts model
is often referred as `weak first-order transition', because the
latent heat $L \sim 0.0265$ is especially small compared with
the cases $q \ge 6$.~\cite{Wu} To distinguish the weak
first-order transition from the second-order transition is a
remaining problem in numerical analyses of statistical models,
and therefore tow-dimensional $q = 5$ Potts model has been
numerically investigated.~\cite{GWilson,Yamagata,Janke} At
present, it is possible to detect non second-order tendencies in
the order parameter~\cite{Janke} or the specific
heat.~\cite{Yamagata} However, quantitative observation of the
latent heat $L$ has not been performed. In this short note we
calculate the latent heat $L$ of the $q = 5$ Potts model
directly by subtracting the site energy of the ordered ($=$
ferromagnetic) phase at the transition temperature from that of
the disordered ($=$ paramagnetic) phase. For this purpose we use
the corner transfer matrix renormalization group (CTMRG)
method.~\cite{CTMRG1,CTMRG2,CTMRG3,Escorial}

The outline of CTMRG is as follows. Let us consider a square
cluster of $q$-state Potts model whose linear size is $N$. The
cluster consists of four sub-clusters of the linear size
$(N+1)/2$, which are called as `corners'.~\cite{Baxter} The
partition function $Z_N$ is calculated as the trace of $\rho_N =
\left( A_{(N+1)/2} \right)^4$, where $A_{(N+1)/2}$ is so called
`corner transfer matrix' (CTM) which transfers column spins into
row spins.~\cite{Baxter} The point of CTMRG is that $\rho_N$ can
be regarded as a kind of density matrix in DMRG, because its
trace is the the partition function $Z_N$. We can therefore
apply the renormalization procedure to
CTM,~\cite{CTMRG3,Escorial} and transform $A_{(N+1)/2}$ into a
renormalized one ${\bar A}_{(N+1)/2}$, which is an
$m$-dimensional matrix. Trace of ${\bar \rho}_N = \left( {\bar
A}_{(N+1)/2} \right)^4$ gives a precise lower-bound ${\bar Z}_N$
for the exact partition function $Z_N$, where the difference
$Z_N - {\bar Z}_N$ approaches to zero with increasing $m$. We
can increase the linear size of the corner one by one using a
recursive relation between ${\bar A}_n$ and ${\bar A}_{n+1}$,
and obtain ${\bar A}_{(N+1)/2}$ for arbitrary $N$. The
renormalized CTM ${\bar A}_{(N+1)/2}$ thus obtained successively
gives the approximate partition function ${\bar Z}_N$. In
addition to the partition function, we can calculate the site
energy and the order parameter at the center of the square
cluster.

The partition function of the two-dimensional $q = 5$ Potts
model is given by
\begin{equation}
Z^N = \sum_{\{\sigma\}} {\rm exp}
\bigg\{ K \sum_{\langle ij\rangle} \delta(\sigma_i,\sigma_j)
\bigg\},
\end{equation}
where $N$ is the linear dimension of the finite size system,
$\sigma_i$ is the five-state spin variable on the lattice point
$i$,  $\langle ij\rangle$ specifies the neighboring spin pairs,
and $\delta(\sigma_i,\sigma_j)$ is equal to unity only when
$\sigma_i = \sigma_j$ and zero otherwise. We consider both
paramagnetic boundary conditions (PBC), where arbitrary spin
configurations are allowed at the boundary, and ferromagnetic
boundary conditions (FBC), where boundary spins are fixed to the
same direction. In the limit of $N \rightarrow \infty$, the $q =
5$ Potts model shows first-order phase transition when the
parameter $K$ is equal to
\begin{equation}
K_c = {\rm ln} \, (\sqrt{5} + 1) \, .
\end{equation}
At the transition point, the site energy
\begin{equation}
U = 2 \langle \delta({\sigma}_i, {\sigma}_j) \rangle
\end{equation}
of the disordered ($=$ high-temperature) phase $U^{+}$ differs
from that of the ordered ($=$ low-temperature) phase $U^{-}$ by
the latent heat $L \sim 0.0256$. The average of $U^{+}$ and
$U^{-}$ is simply expressed as $U_0 = 1/2 + 1/2\sqrt{5} $, that
is obtained by the duality relation.~\cite{Baxter,Wu}

\begin{figure}
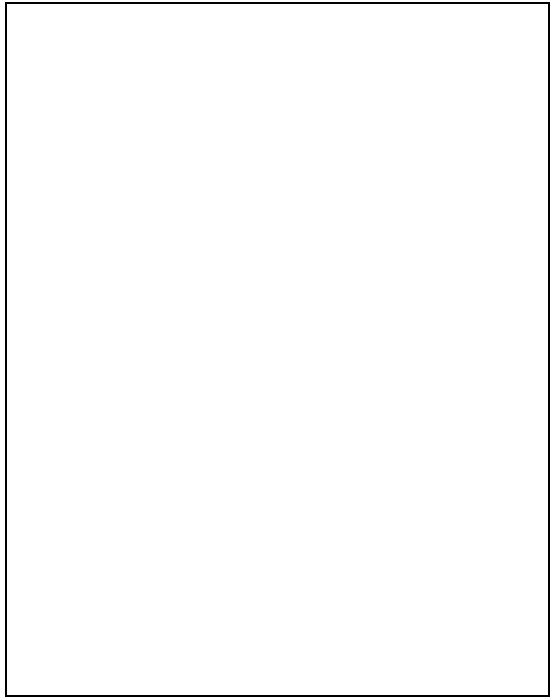

\figureheight{9cm}
\caption{
System size dependence of the site energy for both fixed and
free boundary conditions.
}
\label{fig:1}
\end{figure}

In order to estimate $U^{\pm}$ numerically, we first calculate
the site energy $U^{+}_N(m)$ and $U^{-}_N(m)$ at the center of
square cluster of size $N$, respectively, by imposing PBC and
FBC when $K = K_c$. The number $m$  is the matrix dimension of
the renormalized CTM ${\bar A}_{(N+1)/2}$.  It should be noted
that $U^{\pm}$ is equal to ${\rm lim}_{m \rightarrow \infty}
{\rm lim}_{N \rightarrow \infty} U^{\pm}_{N}(m)$. Figure 1 shows
$U^{\pm}_N(m)$ for $m = 40, 67$ and $200$ up to $N = 4000$. We
choose $U_0$ as the zero of the site energy. The $m$ dependence
of $U^{\pm}_N(m)$ is not negligible in the large $N$ region,
where the energy $U^{\pm}_N(m)$ approaches to
$U^{\pm}_{\infty}(m)$ exponentially with respect to $N$.

\begin{figure}
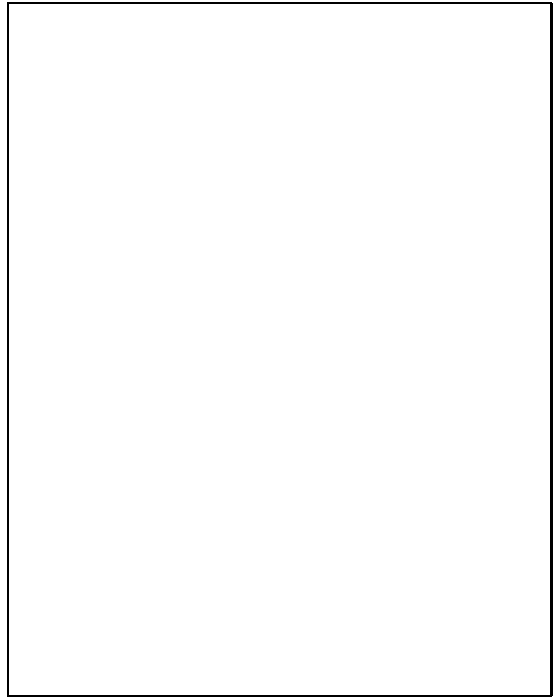

\figureheight{9cm}
\caption{
The $m$ dependence of $U^{\pm}_{\infty}(m)$. The $1 / m$
extrapolation  gives the latent heat $L$.
}
\label{fig:2}
\end{figure}

Figure 2 shows the $m$ dependence of $U^{\pm}_{\infty}(m)$.
There is $m$ dependence because the energy cut-off introduced by
the renormalization group transformation is comparable to the
latent heat $L$. As shown in Fig.2, $U^{\pm}_{\infty}(m)$ is
nearly linear in $1/m$, and  therefore we may linearly
extrapolate $U^{\pm}_{\infty}(m)$ to obtain $U^{\pm} =
U^{\pm}_{\infty}(\infty)$. As a result we obtain $U^{+} = U_0 -
0.014$ and $U^{-} = U_0 + 0.013$, where the obtained $U^{+}$ and
$U^{-}$ are slightly lower than the exact ones. We finally
obtain the latent heat $L = U^{-} - U^{+} = 0.027$, which
quantitatively agrees with the exact one $L \sim 0.0265$. As far
as we know, this is the first quantitative estimate of the
latent heat by numerical calculation.

The authors would like to express their sincere thanks to
Y.~Akutsu, Y.~Kato, M.~Kikuchi, H.~Otsuka, and A.~Yamagata for
valuable discussions about the weak first-order phase
transition. K.~O. is supported by JSPS Research Fellowships for
Young Scientists. The present work is partially supported by a
Grant-in-Aid from Ministry of Education, Science and Culture of
Japan. Most of the numerical calculations were done by NEC
SX-3/14R and SX-4 in computer center of Osaka University.

\end{document}